# Инновационные формы продаж в электронной коммерции
М.Л. Калужский

*Аннотация*: Статья о трансформации дистанционных продаж в условиях электронной коммерции. Автор классифицирует и анализирует особенности новых методов электронных продаж в виртуальной среде.

*Ключевые слова*: электронная коммерция, интернет-маркетинг, интернет-продажи, интернет-аукцион.

# Innovative forms of sales in the e-commerce
M.L. Kaluzhsky

*Abstract*: Article about transformation of methods of remote sales in the conditions of e-commerce. The author classifies and analyzes features of new methods of electronic sales in the virtual environment.

*Keywords*: e-commerce, Internet-marketing, Internet-sales, Internet-auction.

Доминирующая сегодня в электронной коммерции распределительная концепция маркетинга неразрывно связана с процессом формирования новых форм и методов организации продаж. В традиционном маркетинге основная идея этой концепции заключается в том, что эффективность продаж находится в прямой зависимости от оптимальности распределения товара по территории рынка.[1] Применительно к интернет-маркетингу распределительная концепция подразумевает максимизацию доступности товарных предложений для потенциальных покупателей в сети Интернет.

Важной особенностью электронной коммерции является то, что источником маркетинговой деятельности в Интернете может выступать любой из трёх типов участников процесса товародвижения: производители, посредники и покупатели. Это предполагает применение широчайшего спектра инновационных форм и методов организации продаж, список которых постоянно пополняется за счёт включения новых целевых рынков и аудиторий.

В электронной коммерции такой подход подразумевает использование специфически виртуальных каналов сбыта и коммуникации:

**I. Интернет-магазины** – специализированные сайты, в автоматическом режиме торгующие товарами от имени владельца. Интернет-магазины были первой формой электронной коммерции в сети Интернет. Они прошли эволюцию от примитивных прайс-листов с изображениями товаров для сложных торговых систем автоматизированным приёмом платежей и обработкой заказов. Их отличительной особенностью является полный цикл торговых услуг, оказываемых покупателям: от приёма заказов до отгрузки продукции.

По уровню организации продаж можно выделить три основных вида интернет-магазинов:

1. *Интернет-витрины*, на которых представлено описание товаров, условия их продажи и поставки, а также контактные реквизиты продавца.[2] Их трудно назвать полноценными интернет-магазинами, но основные атрибуты магазинов (товары, продавец и место продаж) здесь присутствуют.

До сих пор подавляющая часть сайтов российских компаний представляет собой такие сайты-визитки. Эффективность охвата аудитории ими крайне невелика, ассортимент продукции и объёмы заказов минимальны. Основная маркетинговая функция, которую они выполняют, заключается в расширении коммуникативных возможностей.

2. *Самостоятельные интернет-магазины* с регистрацией покупателей, возможностью интерактивного отбора товаров и автоматизированной оплатой через электронные платёж-ные системы. Отличительной чертой таких магазинов является усечённость виртуальных возможностей – только продажи происходят в Интернете. Управление товарными запасами и складской инфраструктурой происходит как в традиционной торговле.

---

[1] Калужский М.Л. Практический маркетинг. – СПб.: Питер, 2012. – С. 21.
[2] Юрасов А.В. Электронная коммерция. – М.: Дело, 2003. – С. 131-132.

Недостатком самостоятельных интернет-магазинов является их зависимость от поисковых систем для привлечения новых покупателей и от непрерывно сокращающегося числа старых клиентов. Многие из старейших российских интернет-магазинов (eHouse, Утконос и др.) испытывают сегодня серьёзные финансовые проблемы, не поспевая за стремительным изменением маркетинговых технологий в сети.[3]

3. *Интегрированные интернет-магазины* являются составной частью крупных ритейлинговых сетей. Это наиболее конкурентоспособный вид интернет-магазинов, обладающий целым рядом преимуществ: высоким уровнем доверия покупателей, разветвлённой сетью пунктов выдачи товара и слабой зависимостью от поставщиков.

Снижение продаж просто вынуждает ритейлеров ускоренно осваивать технологии электронных продаж. Сегодня такие интегрированные интернет-магазины стабильно занимают первые места в рейтингах интернет-посещений.[4] Однако в перспективе ритейлеры всё же проигрывают виртуальным компаниям за счёт больших товарных запасов, медленного обновления ассортимента и нарастающих проблем с поставщиками.

В настоящее время общая тенденция развития электронной коммерции свидетельствует о завершении процесса вытеснения фирменных интернет-ресурсов с лидирующих позиций на рынке. Никакие SEO-оптимизации контента уже не способны поднять уровень конкурентоспособности интернет-ресурса, если его ассортимент ограничен собственной продукцией и услугами продавца, а интернет-торговля является разновидностью рекламной кампании.

Так, например, из ста лидеров российской электронной коммерции по версии ООО «Рамблер Интернет Холдинг» только 9 сайтов принадлежит компаниям, использующим Интернет в качестве канала маркетинговых коммуникаций.[5] На их долю приходится 2,06% (около 100 тыс. человек) от общего количества уникальных посетителей в сутки.[6]

Самостоятельные интернет-магазины, использующие Интернет в качестве основного канала сбыта, согласно проведённому анализу, представлены в «ТОП-100: торговля» Рамблера 24-ю интернет-ресурсами. Это составило 11,02% (около 538 тыс. человек) от общего количества уникальных посетителей. Средний показатель посещаемости составил 22 тыс. человек на один сайт в сутки при том, что в предыдущем случае этот показатель составил около 11 тыс. человек.

И, наконец, подразделения крупных ритейлинговых сетей оказались представленными лишь 9-ю интернет-ресурсами. Что составило 11,77% (около 574 тыс. человек) от общего количества уникальных посетителей со средним показателем 63,8 тыс. уникальных посетителей в сутки при пересчёте на один ресурс. Данный сегмент весьма неоднороден, о чём свидетельствует разброс показателей от 440 тыс. уникальных посетителей у тройки лидеров («DNS», «М-Видео», «Эльдорадо») до 26 тыс. уникальных посетителей у аутсайдеров («Триал-Спорт», Аптека «ИФК», фото и видео-техника «АВС»).

Следует отметить и тот факт, что ритейлинговые компании, лидирующие в интернет-продажах, лидируют и в традиционной розничной торговле. Всех их объединяет одно – более выгодное для покупателей ценообразование, чем у конкурентов. Это ещё раз подтверждает тезис о приоритетном значении цены в интернет-маркетинге для покупателей (после распределения).

Таким образом, можно уверенно заявлять о том, что институционализация традиционной магазинной торговли в сети Интернет близка к своему логическому завершению. Наибольшим успехом здесь пользуются интернет-магазины крупных розничных торговых сетей, предлагающие относительно низкие цены и максимальный сервис. Применительно к ритейлерам уникальность сервиса заключается в том, что покупатель может быстро получить товар в местном магазине торговой сети.

---

[3] EHouse не выдержал 003. – http://www.oborot.ru/article/422/24.
[4] Rambler. Топ 100. Торговля. – http://top100.rambler.ru/navi/?theme=673.
[5] Там же.
[6] Исследование проводилось методом моментных наблюдений 18.08.2012 на основании данных интернет-ресурса «Rambler» – http://www.rambler.ru.

Однако даже при условии быстрого получения товара низкая цена предопределяет, но не гарантирует маркетинговый успех продавца. Проблема растущей конкуренции со стороны виртуальных компаний заключается сегодня в том, что они вполне могут конкурировать с ритейлерами в ценообразовании, а в скорости обновления ассортимента они вообще вне конкуренции.

Фактически ритейлеры оказались между двух огней. С одной стороны, издержки в розничной торговле мешают им снижать цены до необходимого уровня. С другой стороны, попытки ритейлеров устанавливать различные цены в розничной и интернет-торговле ведут к оттоку покупателей из розницы в Интернет. В условиях экономического кризиса покупатели сначала изучают наличный товар в розничной торговле, а затем заказывают его по более дешёвой цене в интернет-магазинах.[7]

В качестве частичного решения проблемы некоторые торговые сети применяют стратегию «сбытовой дискриминации» покупателей. Эта стратегия выражается в том, что ритейлеры открывают одновременно два интернет-ресурса – под старой и новой торговой маркой. Ресурс под старой маркой выполняет функцию интернет-витрины, где цены и ассортимент рассчитан на традиционных покупателей, для которых наличие товара на сайте является поводом посетить магазин.

Ресурс под новой торговой маркой рассчитан на покупателей, для которых цена имеет решающее значение. Ранее на этом рынке доминировали полностью или частично виртуальные компании, не обладающие торговыми площадями и связанными с ними издержками. Интернет-магазины ритейлеров при равных с ними ценах обладают огромным преимуществом ввиду бесплатной доставки товаров покупателям.

В качестве примеров такой маркетинговой политики можно привести, например, интернет-магазин «RBT»[8] торговой сети «Эксперт» или электронный дискаунтер «TechnoPoint»[9] торговой сети «DNS». Оба этих интернет-магазина никак не ассоциируют себя с головной компанией и предназначены для обслуживания оппортунистически настроенных по отношению к ней покупателей. При этом используется типичный для виртуальных компаний принцип «торговли со склада» с прямой бесплатной доставкой товара покупателю.

Интернет-магазинами исчерпывается список традиционной торговли в Интернете. Все остальные участники электронной коммерции не обладают полным циклом торговых услуг и являются либо виртуальными посредниками, либо виртуальными покупателями.

**II. Интернет-аукционы** – электронные ресурсы, предоставляющие пользователям возможность покупать и продавать товары на условиях аукционных торгов. Если за пределами Интернета аукционы сохранились в основном в торговле эксклюзивными товарами и организации государственных (корпоративных) закупок, то в Интернете эта форма электронной коммерции процветает.

Основное преимущество интернет-аукционов перед обычными аукционами заключается в их большей открытости и доступности для торговли нетрадиционными аукционными товарами. Традиционные аукционы подразумевают наличие уникальности и неповторимости товара, имея своей целью конкурентную продажу товара по максимальной цене узкой целевой аудитории. В последние годы даже старейшие аукционные дома Европы вводят возможность виртуального участия в торгах, что, однако, не меняет общей закономерности.[10]

Основная масса товаров на интернет-аукционах, наоборот, не обладает ни уникальностью, ни неповторимостью. Сюда попадают товары, которые не подходят для обычных аукционов. Главная их цель заключается не в том, чтобы получить комиссионное вознаграж-ние за счёт продажи эксклюзивных товаров по эксклюзивной цене, а в том, чтобы получить прибыль за счёт аукционных сборов с большого числа автоматизированных продаж. Основным источником прибыли здесь являются объёмы продаж, а не эксклюзивность товаров.

---

[7] Юрасов А.В. Электронная коммерция. – С. 134.
[8] Интернет-магазин бытовой техники «RBT» – www.rbt.ru.
[9] Электронный дискаунтер «TechnoPoint» – http://ekaterinburg.technopoint.ru.
[10] См., напр.: Сайт аукционного дома «Sotheby's». – http://www.sothebys.com.

Так, если традиционный аукцион подразумевает наличие товара у аукционера и обязательность его экспертизы, то интернет-аукционы лишь предоставляют место и инструменты участникам электронной коммерции для совершения онлайн-торгов. Это особый тип посредников в сети Интернет, который, как и интернет-магазины, делится на три основные категории:

1. *Скандинавские интернет-аукционы* – представляют собой вид интернет-аукциона на повышение цены с оплатой за каждую ставку. В некоторых странах (например, в Италии) они приравниваются к азартным играм и требуют наличия лицензии на организацию игорной деятельности. В России статус скандинавских аукционов законодательно пока не определён.

Система торгов заключается в том, что участники торгов платят за право делать ставки – за каждый шаг аукциона в отдельности. Начальная ставка всегда очень низка, а результаты торгов отображаются на сайте интернет-аукциона в реальном времени. После очередной ставки торги продляются на определенное время (обычно несколько минут). Выигрывает тот, чья ставка была последняя до истечения контрольного времени аукциона.

Система скандинавских аукциона была впервые разработана ныне обанкротившейся германской компанией «Entertainment Shopping AG» в 2005 году. Наибольшее развитие она получила в США и Европе. Сегодня в США действуют более сотни «Penny Auction» с ежегодным оборотом около 2 млрд. долларов.[11] Американские держатели скандинавских аукционов объединены в две ассоциации: «Entertainment auctions association»[12] (EAA) и «Penny Auctions Merchants Association»[13] (PAMA).

В России скандинавские аукционы также достаточно широко представлены. Отчасти это связано с тем, что одним из мировых лидеров разработки программного обеспечения для аукционной торговли является белорусская компания «SoftSwiss», поставляющая «под ключ» интернет-решения на русском языке.[14]

Старейший в России скандинавский аукцион «Gagen» существует с 2008 года.[15] Отличительной чертой отечественных «скандинавских» аукционов является наличие в их структуре интернет-магазинов, где можно без торгов приобрести продаваемые товары. Судя по рейтингу «Rambler. Топ 100. Торговля» существенной роли в развитии электронной коммерции они не играют.[16]

2. *Обратные интернет-аукционы* – представляют собой вид аукциона покупателя на понижение стартовой цены. На обратном аукционе покупатель сам устанавливает неизвестную продавцам минимальную цену покупки, а продавцы соревнуются – кто быстрее достигнет этой цены. Тут возможны два варианта: либо торги завершаются по достижении желаемой цены продавца, либо время торгов фиксировано, но сделка состоится только при достижении желаемой цены покупателя.[17]

В качестве наиболее успешного зарубежного примера обратного интернет-аукциона можно привести американскую компанию «Priceline.com Inc.».[18] Аукционом это можно назвать с известной долей условности, так как речь идёт скорее о посредничестве. Компания специализируется на организации покупок авиабилетов, аренде автомобилей, туристическом и гостиничном бизнесе.

---

[11] В Европе и США скандинавские аукционы называются «Penny Auction», т.е. копеечные аукционы. К Скандинавии они никакого отношения не имеют.
[12] Сайт ассоциации «Entertainment auctions association». – http://entertainmentauctionassociation.com.
[13] Сайт ассоциации «Penny Auctions Merchants Association». – http://www.pennyauctionassociation.org.
[14] См.: Сайт компании «SoftSwiss». – http://softswiss.com.
[15] См.: Сайт Системы скандинавских аукционов «Gagen» (Россия). – http://gagen.ru.
[16] Rambler. Топ 100. Торговля. – http://top100.rambler.ru/navi/?theme=673.
[17] Юрасов А.В. Электронная коммерция. – С. 177-178.
[18] Сайт компании «Priceline.com Inc.». – http://www.priceline.com.

Покупатель здесь заранее устанавливает максимальную цену, которую он готов заплатить за предоставляемые услуги. Затем поставщики услуг (гостиницы, авиакомпании и т.п.) заполняют «горящие» места за счёт лучших из поданных предложений. При этом покупатели не могут самостоятельно определять точное время оказания услуг.

В качестве наиболее типичного российского аналога обратных аукционов можно привести Портал государственных закупок РФ.[19] Портал действует по схеме «G2B» (Government to Business). Государственные органы размещают условия закупок, а участники аукционных торгов соревнуются, предлагая наиболее выгодные варианты.

Сюда же можно отнести голландские аукционы, основанные на торгах с постепенным понижением цены. Торги начинаются после внесения регистрационного взноса определенным заранее числом участников. Аукционер (реальный или виртуальный) постепенно понижает цену, и товар достаётся тому, кто первый согласился купить товар по текущей цене. Остальные участники остаются без товара и без регистрационного взноса.

В России голландские аукционы проводит, например, аукционный дом «Альянс».[20] Если не брать во внимание госзакупки, то существенной роли в развитии рыночной электронной коммерции обратные аукционы пока не играют.

3. *Классические интернет-аукционы* – напоминают обычные аукционные торги на повышение цены. Покупатели самостоятельно «выставляют» лоты на торги через заполнение автоматизированных форм на сайте аукциона. Далее торги происходят в автоматическом круглосуточно режиме. Длительность торгов, условия поставки и оплаты определяет продавец в соответствии с правилами аукциона.

Нарушение этих правил любым из участников торгов ведёт к приостановке или аннулированию пользовательского аккаунта. Многие аукционы предусматривают возможность компенсации убытков покупателя в случае мошеннических действий продавца. Например, программа защиты покупателей на интернет-аукционе «Молоток» предусматривает компенсацию в сумме до 5.000 рублей в случае непоставки товара или мошеннических действий со стороны продавца.[21]

Преимущество классических интернет-аукционов среди других форм электронной коммерции заключается в повышенной доступности торгов для участников. Любой желающий может свободно выставлять товар на торги или покупать выставленный товар. Классические аукционы объединили в себе блошиный рынок, комиссионный магазин и доску объявлений.

Первый и наиболее успешный классический аукцион «eBay» был запущен в 1995 году в США. По данным 2011 года число его постоянных участников достигло 100 млн. человек, а общая сумма годовых продаж достигла 68,6 млрд. долларов.[22]

В России таким лидером стал интернет-аукцион «Молоток» («Aukro»), запущенный в декабре 1999 года. В 2011 году объём продаж Молотка составил 2,7 млрд. рублей, со средней посещаемостью 200 тыс. уникальных посетителей в сутки.

Трансформация классических интернет-аукционов включает в себя несколько стадий.

На первой стадии интернет-аукцион напоминает блошиный рынок и отличается от электронной доски объявлений лишь системой рейтингов и несколько большей защищенностью покупателей. С учётом отсутствия комиссии за выставление лотов и очень низкой комиссии с продаж это привлекает большое количество участников.

На второй стадии интернет-аукцион вводит плату за выставление лота и повышает сборы с продаж. Это отсеивает т.н. «мусорных продавцов» с большим количеством неликвидных лотов. Например, на «Молоток.Ру» в некоторых категориях до 80% листингов еще недавно составляли неликвидные лоты от 2% продавцов. Вместе с тем, из около 60 тыс. про-

---

[19] Официальный сайт Российской Федерации для размещения информации о размещении заказов. – http://zakupki.gov.ru.
[20] Сайт Аукционного дома «Альянс». – http://adalyance.ru.
[21] Сайт ООО «е-коммерс груп». – http://molotok.ru/country_pages/168/0/education/pok/index.php
[22] Сайт компании «eBay Inc.». – http://www.ebayinc.com/who.

давцов на Молотке примерно 6 тыс. – профессиональные продавцы, которые обеспечивают 85% прибыли аукциона.[23]

Третья стадия связана с институциональным переходом от классического интернет-аукциона к торговой интернет-площадке. Однако интернет-площадка – это уже совсем иная форма электронной коммерции с иными отличительными характеристиками.

**III. Торговые интернет-площадки** – электронные ресурсы, предоставляющие профессиональным продавцам и (или) покупателям программные инструменты и виртуальное пространство для проведения торгов. Основное преимущество интернет-площадок перед другими формами электронной коммерции заключается в большей посещаемости и высоком уровне сервиса, как для продавцов, так и для покупателей. По происхождению наиболее успешные торговые площадки делятся на четыре основные категории:

1. *Выросшие из интернет-магазинов*. К первой категории можно отнести одну из крупнейших в мире интернет-площадок «Amazon», созданную в 1995 году в США как крупнейший в мире виртуальный книжный интернет-магазин.[24] Однако затем ассортимент был расширен за счёт других товарных групп и к торговле допустили независимых продавцов. По информативности и сервисности «Amazon» обладает всеми преимуществами интернет-аукционов за исключением конкурентных торгов. Не случайно по итогам 2011 года чистые продажи «Amazon» составили 48,08 млрд. долларов, в сравнении с 34,2 млрд. долларов в 2010 году.[25]

В России аналогом «Amazon» является интернет-магазин «Ozon», также начавший свою деятельность в 1998 году как виртуальный книжный интернет-магазин.[26] Несмотря на существенно меньшие показатели продаж и более узкий ассортимент, «Ozon» является безусловным лидером в российской электронной коммерции. Так, в 2011 году оборот «Ozon» достиг 8,87 млрд. рублей (рост на 78% по сравнению с 2010 годов), а число покупок составило 2,42 млн.[27]

Слабое место «Ozon» заключается в том, что компания не вышла за формальные рамки интернет-магазина. Хотя на сайте присутствует большой ассортимент товаров, начиная от туристических путёвок и заканчивая антиквариатом. Однако «Ozon» до сих пор торгует от своего имени, предпочитая не допускать к прямым торгам сторонних продавцов. Следствием этого является потеря потенциальных покупателей и упущенные возможности роста. В то же время безусловно сильной стороной «Ozon» является налаженная система транспортной логистики, опирающаяся на пункты выдачи товара во всех крупных городах России.

2. *Выросшие из интернет-аукционов*. Интернет-аукционы становятся торговыми площадками после того, как вводят платный магазинный сервис для профессиональных продавцов. Признаками торговых площадок также являются интегрированные платёжные системы, большая доля продаж товара по фиксированным ценам и ценовая дискриминация непрофессиональных продавцов. Такие торговые площадки обладают гораздо большей конкурентоспособностью в сравнении с интернет-магазинами в силу их отказа от функций ценообразования. Они не продают товары, их продукт – виртуальная среда для совершения сделок и платные сервисы (торговые, платёжные и др.).

Безусловным мировым лидером в этой категории является феноменально успешная торговая площадка «eBay», основанная в 1995 году. На её долю приходится более 100 млн. активных пользователей по всему миру. Оборот «eBay» составил в 2011 году 68,6 млрд. долларов.[28] В 2011 году «eBay» завершил структурный переход от Интернет аукциона к торго-вой площадке. Это выразилось в резком повышении стоимости услуг для торговцев и оттоке мелких продавцов.

---

[23] По материалам Сайта ООО «е-коммерс груп». – http://molotok.ru.
[24] Сайт компании «Amazon Global Resources». – http://www.amazon.com.
[25] Матвеева А. Kindle не вывез Amazon. // Газета.Ру. – 01.02.2012. – http://www.gazeta.ru/financial/2012/02/01/3981917.shtml.
[26] См.: Онлайн-мегамаркет «OZON.ru». – http://www.ozon.ru.
[27] По данным сайта «Ozon»: http://www.ozon.ru/context/detail/id/4504639.
[28] По данным сайта «eBay». – http://www.ebayinc.com/who.

Для примера можно отметить, что при месячном обороте в 500 долларов стоимость услуг «eBay» составляет около 150 долларов или 30%. Процентное соотношение оплаты услуг «eBay» и получаемой прибыли снижается для торговцев с большим оборотом. Повышение тарифов привело к резкому росту прибыли «eBay» (на 400% в 2008 году).[29] Часть продавцов перешла на более мелкие аукционы (например, на европейский «Delcampe»). Однако вскоре началось их возвращение, так как покупатели остались привержены «eBay». Большую роль в успехе «eBay» играет принадлежащий компании крупнейший в мире платёжный сервис «PayPal», число пользователей которого также превышает 100 млн. человек.[30]

В России аналогом торговой площадки «eBay» является интернет-портал «Молоток», входящий в международную группу «Aukro» и позиционирующий себя с недавнего времени как «открытая торговая площадка».[31] Переход от интернет-аукциона к формату торговой площадки в 2012 году выразился в увеличении сборов с продаж с 3% до 15% (товары для взрослых), введении статуса VIP-продавца, персональных страниц магазинов юридических лиц (включая логотип и брендирование). В ближайшее время «Молоток» планирует интегрировать в торговую систему платёжный сервис «PayU», что окончательно утвердит его статус торговой площадки и откроет новые возможности для продавцов.[32]

3. *Независимые торговые площадки*. Такие площадки наиболее успешны в секторе «B2B» при отсутствии развитой торговой инфраструктуры. Непременным условием их формирования является избыток производственных мощностей, наличие большого потенциального спроса на производимую продукцию и отставание в развитии оптового звена. Описанная ситуация особенно характерна для экономики Китая и именно там можно обнаружить наиболее успешные независимые торговые площадки.

Ведущим оператором торговых площадок в Китае является группа компаний «Alibaba Group Holding Ltd», основанная в 1999 году как площадка «B2B» для мелких и средних предприятий, а онлайн продаж. В её состав сегодня входят:
– платёжная система «Alipay» (600 млн. пользователей),[33]
– электронная оптовая торговая площадка («B2B») «Alibaba.com»,[34]
– онлайн-рынок розничной торговли «Taobao»[35] с международным подразделением «Aliexpress»,[36]
– компания по разработке и продаже готовых решений для управления коммерческой деятельностью в сети Интернет «Alisoft»,[37]
– специализированный сайт по обмену интернет-рекламой между веб-издателями и рекламодателями «Alimama» и др.

Всего только на «Alibaba.com» (по данным сайта) зарегистрировано 72,8 млн. пользователей из 240 стран мира. Продажи «Alibaba.com» составили в 2011 году 6,4 млрд. юаней, увеличившись по сравнению с 2010 г. на 15,5%.[38] При этом сайт не взимает плату за совершение сделок и не требует процент с продаж, ограничиваясь ежегодными членскими взносами (т.н. «Gold Supplier»). Следует отметить, что в последние году наблюдается снижение числа платных аккаунтов на «Alibaba.com» из-за высокого размера этих взносов (2.999 долларов в год).

---

[29] По данным сайта «РБК»: http://top.rbc.ru/economics/22/01/2009/275613.shtml.
[30] Сайт электронной платежной системы «PayPal». – https://www.paypal.com.
[31] Открытая торговая площадка «Молоток». – http://molotok.ru.
[32] Готовимся к запуску PayU. // Новости Молоток.Ру. – 06.07.2012. – http://molotok.ru/News.php.
[33] Сайт электронной платежной системы «Alipay». – http://global.alipay.com.
[34] Сайт торговой площадки «Alibaba.com». – http://russian.alibaba.com.
[35] Сайт онлайн-рынка розничной торговли «Taobao». – http://www.taobao.com.
[36] Сайт онлайн-рынка розничной торговли «Aliexpress». – http://aliexpress.com.
[37] Сайт компании «Alisoft». – http://www.alisoft.com.
[38] Квартальная прибыль владельца ведущего в Китае сайта интернет-торговли сократилась на 6% / ИА «Финмаркет». – 21.02.2012. – http://www.finmarket.ru/z/nws/news.asp?id=2765873.

Кроме «Alibaba Group Holding Ltd» на китайском рынке представлено множество других конкурирующих площадок (например, «BuyWholesaleLots.com», «DHGate» и целый ряд др.). Недостаточная развитость традиционной торговой инфраструктуры и относительная закрытость китайской экономики создали благоприятные условия для их бурного развития в условиях промышленного производства.

В России также присутствует достаточно большое число независимых торговых площадок, ориентированных в основном на сектор «B2B». Однако они не играют существенной роли в развитии отечественной электронной коммерции. Причина кроется в том, что для их развития требуется наличие значительного дисбаланса между спросом и предложением в условиях недостаточно развитой торговой инфраструктуры. Поэтому в России сегодня независимые торговые площадки не перспективны.

4. *Зависимые торговые площадки*. Такие площадки создаются в секторе «B2B» крупными покупателями с целью проведения конкурентных торгов в своих интересах. Большие заказы привлекают потенциальных участников, которые параллельно используют коммуникативные возможности торговой площадки для совершения сделок между собой. Отличительная черта зависимых торговых площадок – их узкая специализация, ориентированная на специфику компании-владельца.

Хрестоматийным стал пример американской компании «ChemConnect Inc.». В 2001 году эта компания, используя свои возможности держателя одной из первых интернет-площадок для торговли химикатами и пластмассами, обеспечила на льготных условиях свои потребности в данной продукции.[39]

В России зависимые торговые площадки ориентированы либо на крупные корпоративные, либо на государственные или муниципальные заказы. В качестве примера можно привести, например, «Железнодорожную торговую площадку» ориентированную на нужны ОАО «РЖД».[40]

Другим примером могут стать торговые площадки, предназначенные для размещения заказов в соответствии с Федеральным законом РФ № 94-ФЗ «О размещении заказов на поставки товаров, выполнение работ, оказание услуг для государственных и муниципальных нужд». В соответствии с приказом Минэкономразвития № 428 от 26.10.2009 г. пять электронных площадок было аккредитовано на 5 лет в качестве национальных операторов электронных торгов.

**IV. Сервисы коллективных покупок** – электронные ресурсы, распространяющие информацию о скидках продавцов среди потенциальных покупателей. Это своего рода система коллективных покупок, предоставляющая продавцам возможность гарантированного привлечения большого числа клиентов за счёт предоставления значительных скидок.

Сервис коллективных покупок получает прибыль от продажи прав на получения скидки. При этом скидки делятся на два вида:

1. *Стандартная скидка* – подразумевает, что после накопления до определённого количества предварительных заявок потенциальные покупатели вносят деньги для приобретения товара. Администрация сервиса от имени покупателей приобретает товар по оптовой цене у поставщика и организует его доставку покупателям. Эта форма обычно применяется для формирования круга постоянных клиентов на начальной стадии становления сервисов коллективных покупок.

В России, например, такой сервис в форме обратного аукциона был запущен компанией «Public Sale Company Ltd.» в 2011 году.[41] Сервис собирает заявки на товары от покупателей, на основе которых продавцы делают покупателям свои ценовые предложения, тем самым получая доступ к целевой аудитории потенциальных покупателей. Кроме того, сервис предоставляет возможность покупателям объединяться для выставления групповых заявок с целью получения оптовых скидок от продавцов.

---

[39] Бергер Э.Дж. Цепи поставок: лучшие из лучших. / Управление цепями поставок. Под ред. Дж.Л. Гатторны. – М.: Инфра-М, 2008. – С. 562.
[40] Сайт «Железнодорожная торговая площадка». – http://www.depo-portal.ru.
[41] Сервис коллективных покупок «Top10.ru». – http://top10.ru.

2. *Купонная скидка* – подразумевает, что покупатели приобретают у сервиса купоны на право получения скидки. Купоны покупателей накапливаются до определённого количества, после чего они активируются и покупатели получают право на скидку. Если необходимого количества купонов собрать не удаётся, то акция считается несостоявшейся, а покупатели получают деньги обратно на счёт в «Личном кабинете».

Купонными сервисами применяется два вида купонов: купоны для обмена на товар и купоны для получения скидки. Оплата услуг купонного сервиса покупателем производится с помощью кредитных карт, электронных денег или любым доступным способом. Затем на почтовый адрес покупателя приходит письмо с купоном для предъявления продавцу товара или услуги.

Владелец купона заинтересован в привлечении дополнительных участников для ускорения активации купонов. Кроме того, купонные сервисы приобретают в лице покупателей бесплатных рекламных агентов, предлагая им бонусы за привлечённых новых участников.

Крупнейшей компанией в России является филиал сервиса коллективных покупок «Groupon» созданного в 2008 году в США группой энтузиастов. Практически сразу после создания сервис «Groupon» продемонстрировал феноменальные показатели роста продаж. Уже в 2009 году «Groupon» продал в США более 4 млн. купонов на сумму около $150 млн. долларов.[42] По итогам 2010 года доходы Groupon составили 312,9 млн. долларов, а в 2011 году выросли на 419%, составив 1,6 млрд. долларов.[43]

В 2010 году «Groupon» выкупил российский купонный сервис «Darberry», преобразовав его в январе 2011 года в «Groupon Russia». По данным маркетингового исследования компании «РосБизнесКонсалтинг» в 2011 году более 55% российских пользователей купонных сервисов приобретали купоны на сайте «Groupon.Ru».[44]

Всего в 2011 году численность зарегистрированных покупателей сервиса «Groupon.Ru» в России составила около 1,5 млн. человек, превысив аналогичные показатели основных конкурентов: «Biglion» и «КупиКупон» в 5 раз, «Выгода» в 7,5 раз, «Big Buzzy» в 10 раз, а «Покупон» в 30 раз.[45] Около 100 других сервисов коллективных покупок занимают незначительную долю рынка, значительно отставая от лидеров. Можно утверждать, что рынок купонных сервисов уже достиг своего максимума и теперь борьба идёт за удержание клиентской базы и за привлечение предлагающих скидки компаний.

Поэтому ведущие купонные сервисы постепенно переходят к практике прямых продаж купонов, отказываясь от фиксированных минимумов продаж. Такая стратегия доступна только лидерам этого рынка. Она возможна только при наличии большого числа посетителей, делающего ненужной принудительную фиксацию объёма предварительных заявок по акциям. С другой стороны, регулярное использование продавцами купонов для привлечения посетителей объективно снижает численность обычных покупателей. Поэтому некоторые продавцы целиком переходят на работу с купонами, сначала неоправданно завышая цены, а затем снижая их до близкого к среднему уровня.

Такая политика вызывает оправданные нарекания покупателей, поскольку заниженные цены не предполагают высокого качества предлагаемых товаров и услуг. В результате дискредитируется сама идея купонных сервисов, начинается отток покупателей и выжить на таком рынке могут только наиболее крупные компании с большим ассортиментом предлагаемых товаров и услуг.

---

[42] Русяева П. С миру по скидке // Коммерсантъ. Секрет Фирмы. – № 9 (301) – 13.09.2010. – С. 23.

[43] Groupon: доходы в 2011 г. выросли на 419%, а убытки составили $350,8 млн. / «CNews.ru» РИА «РосБизнесКонсалтинг». – 09.02.2012. – http://www.cnews.ru/news/line/index.shtml?2012/02/09/476834

[44] Российский рынок сервисов коллективных покупок: купоны на скидки от 50% до 90%. / РИА «РосБизнесКонсалтинг». – С. 29.

[45] По материалам сайта «Groupon.Ru». – http://groupon.ru/groupon-offer.pdf.

Непременным условием эффективного применения купонных сервисов для стимулирования продаж является наличие большой прибыли у продавца, если речь идёт о товарах (например, суши) или избыточность предложения для рынка, если речь идёт об услугах (например, кинотеатров). Очень часто продавцы используют купонные сервисы для продажи некачественных товаров по завышенным ценам, пользуясь психологической зависимостью постоянных покупателей.

**V. Шоурумы**. В традиционном маркетинге шоурумом называется выставочное пространство, оформленное для представления товаров под определённой торговой маркой или определённой компании (от англ. «*Showroom*» – демонстрационный зал). Классическими шоурумами можно считать, например, показ новой коллекции дизайнера одежды или выставочный зал в автомобильном салоне. Обычно комплектация шоурума включает в себя стойку с информацией о компании и товарах, презентационные стеллажи и рабочие места менеджеров. В отдельных случаях шоурум может включать в себя кинозал и другие инструменты продвижения в зависимости от ситуации.

Развитие электронной коммерции придало понятию «шоурум» множество новых смысловых оттенков. Например, в Европе и США шоурумингом (*Showrooming*) стали называть процесс предварительного изучения товаров покупателями для того, чтобы впоследствии купить его по более дешёвой цене в интернет-магазине.[46] Такой шоуруминг постепенно становится серьёзной проблемой для розничной торговли по всему миру в связи с широким распространением мобильных интернет-продаж. Покупатели имеют возможность осмотреть товары в магазине и тут же сделать заказ на него через Интернет по более низкой цене.

Выросшие из электронной коммерции российские шоурумы приобрели местную специфику, переориентировавшись на торговлю одеждой и аксессуарами. Российские интернет-шоурумы можно разделить на несколько категорий:

1. *Псевдо-шоурумы*. Такой шоурум представляет собой комбинацию сайта или группы в социальной сети с арендованным помещением, куда посетители могут попасть по рекомендации или после регистрации на сайте. Большое значение имеет атмосфера таинственности и эксклюзивности, отделяющая товары в шоу-румах от товаров в обычной торговле. Эта форма коммерции напоминает торговлю фарцовщиков во времена СССР, поскольку часто не предусматривает регистрации налоговых органах.

Ориентируются псевдо-шоурумы обычно на товары из Китая, удовлетворяющие как минимум одному их трёх критериев: копирующие модные бренды, соответствующие текущей моде или более дешёвые, чем в традиционной торговле. Например, достаточно крупный шоурум «Shop Daniel» позиционирует себя как поставщик копий брендовых товаров с китайской торговой интернет-площадки «Taobao».[47]

Товары в псевдо-шоурумы приобретаются на общедоступных торговых площадках («eBay», «Aliexpress», «Sammydress» и т.п.). Клиенты привлекаются через рекомендации постоянных покупателей, блоги, форумы и социальные сети. Основную категорию покупателей составляют покупатели, не освоившие технологии самостоятельных покупок через Интернет. С ростом проникновения Интернета в России маркетинговые возможности псевдо-шоурумов будут сокращаться.

2. *Классические шоурумы*. Это достаточно новая форма интернет-коммуникаций, использующая технологию шоуруминга для продвижения товаров. Отличие классических шоурумов заключается в специализации на узкой группе товаров. Основная аудитория – потенциальные посетители специализированных магазинов, испытывающие нехватку времени на хождение по ним.

В качестве примера можно привести сайт «Showrooms» компании «Fashion Media Group Ltd.», позиционирующий себя в качестве «Онлайн-примерочной лучших магазинов

---

[46] Smith A. 3 Ways to Beat «Showrooming» / DailyFinance. – 25.04.2012. – http://www.dailyfinance.com/2012/04/25/3-ways-to-beat-showrooming.

[47] Сайт интернет-магазина «Shop-Daniel». – http://shopdaniel.ru/page_13.html.

Москвы».[48] На сайте представлены товары около 50 модных бутиков Москвы и более 100 брендов мужской и женской одежды. Сайт предлагает модели одежды из московских бутиков и скидку при резервировании товара.

Другим примером может стать сайт «ShowRoom.ru» компании ООО «ShowRoom.ru», позиционирующий себя как «Информационный портал: Музыкальные инструменты, оборудование и технологии для света, звука, музыки, видео и телепроизводства».[49] Основная аудитория – специалисты узкого профиля, связанные по роду деятельности со спецификой сайта. Маркетинговая идея сайта заключается в аккумулировании специализированных коммерческих предложений и привлечении целевой аудитории.

В целом классические шоурумы в Интернете напоминают виртуальную форму каталожной торговли, дополненную возможностью заказа товара. Основными клиентами таких шоурумов выступают продавцы товаров, расширяющие с их помощью охват целевой аудитории потенциальных покупателей. В этом смысле коммерческая эффективность классических шоурумов значительно превышает показатели традиционной рекламы.

3. *Оптовые шоурумы*. Оптовые шоурумы ориентируются на владельцев небольших бутиков, позиционируя себя в качестве эксклюзивных представителей ведущих мировых производителей. Такая ситуация возможна только при продаже эксклюзивных товаров с виртуальной системой ценообразования, когда продавец на безальтернативной основе может позволить себе установить любую цену.

В качестве примера можно привести московский шоурум «Ли-Лу», позиционирующий себя в качестве эксклюзивного владельца прав на продажу франшиз от «15 ведущих мировых марок на территории России, СНГ и в странах Балтии».[50] Принцип работы этого шоурума заключается в продаже франшиз от имени производителей мировых брендов одежды с последующей поставкой продукции.

Перспективы оптовых шоурумов достаточно туманны, поскольку посредничество, связанное с продвижением конкурирующих брендов, вряд ли привлечёт реальных держателей мировых брендов. Модная одежда является едва ли не единственной сферой, где оптовые шоурумы сегодня могут существовать. С другой стороны, проникновение российских интернет-пользователей на зарубежные торговые площадки объективно снижает актуальность такого посредничества.

**VI. Перспективные формы продаж**. Технологическое развитие электронной коммерции в ближайшие годы будет связано с развитием облачных вычислений, повышением проникновения Интернета и внедрением мобильных технологий связи четвертого поколения по стандарту LTE.[51] Появление новых форм продаж в электронной коммерции, по всей видимости, будет обусловлено новыми технологическим возможностями интернет-коммуникаций (т.е. скоростью и доступностью соединений). Среди внедряемых уже сегодня инновационных форм электронной коммерции можно отметить веб-киоски и интерактивное телевидение.

*Веб-киоски* – форма интернет-торговли, основанная на применении интерактивных экранов, расположенных в местах скопления людей. Покупка совершается путём сканирования смартфоном QR-кода под изображением товара. Затем товар доставляется курьером по указанному покупателем адресу. Такая технология уже используется сегодня британской ритейлинговой компанией «Tesco» для продажи товаров в аэропортах и на автобусных остановках.[52]

*Интерактивное телевидение* – разновидность услуг платного телевидения, основанная на системе контроля доступа пользователей. Суть её заключается в том, что пользователь имеет возможность интерактивно взаимодействовать с местным провайдером услуг, в том числе заказывая через него товары и услуги с доставкой на дом. Отличие от обычной интернет-торговли состоит в локальности использования сети. Например, покупатели оставляют предварительные заказы для поставки товара из ближайшего магазина.

---

[48] Сайт «Showrooms». – http://showrooms.ru.
[49] Сайт «ShowRoom.ru». – http://www.showroom.ru.
[50] Сайт шоурума «Ли-Лу». – http://www.li-lu.ru.
[51] LTE (*Long Term Evolution*) – стандарт передачи данных со скоростью свыше 300 Мбит/сек.
[52] Neville S. Tesco launches UK's first virtual supermarket at Gatwick airport. // Guardian. – 2012. –August, 7.

В любом случае взаимодействие поставщика и покупателя в электронной коммерции представляет собой улицу с двухсторонним движением. Не только продавцы, но и покупатели получают ощутимую выгоду от торговли без посредников, перераспределяя между собой прибыль от трансформации оптово-розничного звена. Не случайно важное преимущество электронной коммерции состоит в широчайшем спектре маркетинговых возможностей, одинаково доступных как для опытных участников рынка, так и для новичков-энтузиастов.